# Short-chain polymer rigidity due to the Debye process of monohydroxy alcohols

C. Gainaru,[1*] R. Figuli,[2] T. Hecksher,[3] B. Jakobsen,[3] J. C. Dyre,[3] M. Wilhelm,[2] R. Böhmer[1]

[1] *Fakultät für Physik, Technische Universität Dortmund, 44221 Dortmund, Germany*
[2] *Karlsruher Institut für Technologie, 76128 Karlsruhe, Germany*
[3] *DNRF Centre "Glass and Time", IMFUFA, Department of Sciences, Roskilde University, Postbox 260, DK-4000 Roskilde, Denmark*

In addition to the ubiquitous structural relaxation of viscous supercooled liquids, monohydroxy alcohols and several other hydrogen-bonded systems display a strong single-exponential electrical low-frequency absorption. So far, this so-called Debye process could be observed only using dielectric techniques. Exploiting a combination of broad-band and high-resolution rheology experiments for three isomeric octanols, unambiguous mechanical evidence for the Debye process is found. Its spectral signature is similar to the viscoelastic fingerprint of small-chain polymers, enabling us to estimate the effective molecular weight for the supramolecular structure formed by the studied monohydroxy alcohols. This finding opens the venue for the application of further non-dielectric techniques directed at unraveling the microscopic nature of the Debye process and for an understanding of this phenomenon in terms of polymer concepts.

* catalin.gainaru@uni-dortmund.de

Several hydrogen-bonded liquids including water, the matrix of life, and important solvents such as the monohydroxy alcohols (MA) ethanol, propanol, etc., display a very strong electrical absorption. This phenomenon, often exploited today, e.g., in microwave ovens, captivated many scientists including Peter Debye who already in 1913 developed one of the first theories for the quantitative description of these highly polar liquids [1]. MAs have long been conceived to form chain-like supramolecular structures [2], a notion backed up by more recent experimental and simulation evidence [3]. Consequently, well-established concepts from polymer theory [4] that predict a distribution of chain lengths were applied to MAs more than 50 years ago [5]. However, this approach was dismissed later because it implies a distribution of molecular relaxation times, at variance with experimental facts [6]. Nevertheless, it is remarkable that up to now features such as viscoelastic "normal modes", which in polymers are coupled to motions of the macromolecule's end-to-end vector [7,8], could not be detected by means of ultrasound, Brillouin scattering, and low-frequency shear experiments [9,10,11]. Hence, mechanical methods, enormously successful when studying the structural ($\alpha$-) relaxation of liquids [8,12], were generally not regarded as useful to study the nature of the typically 10 to 10,000 times slower Debye process. Since the latter could not be detected calorimetrically [13], in line with expectations for polymeric normal modes, most scientists conceive the Debye process as a purely dielectric effect. This view has hampered the microscopic understanding of the Debye process. Recently, nuclear magnetic resonance (NMR) has revealed molecular motions slower than the structural relaxation and, based on a model of transient end-to-end chains formed by hydroxyl groups, has provided clues towards a reconciliation of the Debye shape with the notion of a polymer-like microscopic dynamics [14].

With polymer-based approaches for the description of MAs revitalized, the missing mechanical signature of the Debye process appears more startling than ever. However, as demonstrated below, we have been able to uncover the viscoelastic signature of the Debye process for several MAs as well as to explain the failure of previous attempts to do so. Thus, our work not only identifies the long-sought, missing evidence in the study of MAs, but paves the way for their quantitative description in terms of well-established concepts from polymer science.

Differing widely in their dielectric relaxation strengths [15,16], the isomeric octanols 4-methyl-3-heptanol (4M3H), 5-methyl-3-heptanol (5M3H), and 2-ethyl-1-hexanol (2E1H) (all obtained from Sigma-Aldrich) are studied in this paper. Their dielectric loss spectra, together with that of propylene carbonate (PC) – a typical non-associating liquid devoid of a Debye process [17] – are shown in Fig. 1. For 4M3H the dielectric relaxation strength of the Debye process is smaller than that of the structural relaxation, i.e., $\Delta\varepsilon_D \approx 0.3\Delta\varepsilon_\alpha$ with corresponding dielectric relaxation times $\tau_D \approx 18\tau_\alpha$. For 2E1H the Debye process is much stronger ($\Delta\varepsilon_D \approx 35\Delta\varepsilon_\alpha$) and slower ($\tau_D \approx 700\tau_\alpha$) than the structural relaxation, and for 5M3H an intermediate behavior is found ($\Delta\varepsilon_D \approx 18\Delta\varepsilon_\alpha$, $\tau_D \approx 160\tau_\alpha$) [15,16].

In Fig. 2 we present the complex shear modulus $G^*(\nu) = G'(\nu) + iG''(\nu)$ of 4M3H, covering more than 7 decades in frequency (from $10^{-3}$ to $10^4$ Hz) using a piezo-ceramic transducer technique [18]. A single-peak structure is observed shifting through the frequency window, indicating the temperature dependence of the structural relaxation time. At first glance, the data look unsuspicious, except that a kink appears on the low-frequency flank of the main process. This feature is reminiscent of a terminal mode reported to occur for a variety of short-chain polymers [19,20,21,22]. One recognizes that time temperature superposition is obeyed for $G^*(\nu)$ of 4M3H. Thus, using the peak frequency $\nu_{max}$ as scaling variable, i.e., as effective shift factor, the shear moduli from Fig. 2 can be represented on a reduced frequency scale, see Fig. 3.

Fig. 3 also includes shear data for 4M3H recorded using another spectrometer, the rheometer ARES G2 from TA Instruments, which covers a frequency range from 0.1 to 100 Hz [23]. Fig. 3 demonstrates that the data from the two mechanical techniques are fully compatible. Not only do the spectral shapes agree, but also the resulting shift factors are compatible with the temperature dependent structural relaxation frequencies determined directly from broad-band mechanical measurements; see the Arrhenius plot, Fig. 4. By combining both set-ups, an effective resolution range of about 5 decades is covered, a prerequisite for detecting small mechanical shear moduli.

To compare 4M3H with liquids showing a much stronger dielectric Debye process (like 2E1H) or a weaker one (like 5M3H), or even no Debye process (like PC and several other glass formers [24]), Fig. 3 presents data for those liquids as well. Overall, the main peak of the spectra display a similar shape, but clear differences are recognized at low frequencies. In particular the shear modulus data of 2E1H resemble normal-mode spectra of short-chain polymers [12]. Several polymer-like regimes can be distinguished in Fig. 3 from the frequency-dependent shear moduli written as $G'(\nu) \propto \nu^{-\alpha}$ and $G''(\nu) \propto \nu^{-\beta}$. Generally, the Kramers-Kronig relations imply $\alpha = \beta$; the one exception to this rule, $\alpha = 2$ and $\beta = 1$, marks the terminal Maxwell-mode as observed in the low-frequency limit governed by the steady-flow viscosity. At intermediate frequencies $\alpha \approx \beta \approx 2/3$ resembles the expectation of the bead-spring model including pairwise hydrodynamic interactions [25] and, consistent with this model, $G'' \approx \sqrt{3}\, G'$ holds here, see p. 192 in [26].

For a polymer with DC viscosity $\eta_0$ Rouse theory (see p. 225 in [26]) predicts that the frequency of the lowest mode occurs at $\nu_R = \pi \rho RT/(12 \eta_0 M_{eff})$ which can be identified via the crossover from the terminal to the rubbery regime. With the density $\rho$, the ideal gas constant $R$, the mass $m$ of a 2E1H molecule, and using that $\eta_0 = \lim_{\omega \to 0}(G''/\omega) = 2.2 \times 10^6$ Pa s, an effective molecular weight of $M_{eff} = \nu_\alpha \pi^2 \rho RT/(6 G_\infty \nu_R) \approx 10 m$ is estimated from the data of 2E1H at 160 K [27]. Thus, $N = M_{eff}/m = 10$ can be interpreted as the number of segments (i.e., of 2E1H molecules) within an end-to-end chain of hydroxyl groups. This $N$ is consistent with previous more indirect estimates from dielectric [14] and infrared [28] spectroscopy and with results from simulations [29]. For alcohols in Fig. 3 other than 2E1H, the rubbery regime and thus $N$ is about 3. However, it is reassuring that for the studied alcohols the ratio $\nu_R/\nu_\alpha$ from mechanical spectroscopy displays the same trends as the separation of the two dielectric processes [15,16,28], see Fig. 1.

The contribution of the 'Debye rigidity' to the frequency-dependent shear viscosity is obtained from $\eta'(\nu) = G''(\nu)/(2\pi\nu)$, which follows from the definitions of the dynamic shear viscosity and modulus, in conjunction with the Maxwell relation written here for the α-relaxation as $\eta \approx G_\infty/(2\pi\nu_{max})$. As the inset of Fig. 4 reveals, for 2E1H the viscosity increase $\Delta\eta_D/\eta$ or the rigidity enhancement $\Delta G_D/G_\infty$ due to the Debye process is a factor of ~10. However, the polymer-like effects show up at moduli about 3 orders of magnitude lower than $G_\infty$. This is below the detection limits of typical shear mechanical and ultrasound experiments [11]. The supra-segmental mechanical contribution observed for 2E1H is comparable to that of short-chain polymers [12], but in alcohols it has escaped detection so far because low-modulus rheological studies are rarely performed on liquids and usually not much below ambient temperature. For 4M3H and 5M3H we find $\Delta\eta_D/\eta$ = 2.1 and 3.2, respectively. Including 2E1H as well, these trends reflect those seen in $\Delta\varepsilon_D/\Delta\varepsilon_\alpha$ and $\tau_D/\tau_\alpha$, cf. Fig. 1.

Another striking difference distinguishing the dielectric from the rheological signatures concerns the spectral shape related to the Debye process. On the one hand, in dielectric spectroscopy fast environmental averaging [30] in conjunction with sufficiently rapid chain length fluctuations convey a single-exponential appearance to this process [14,31]. On the other hand, the mechanical rigidity in neat MAs documented above, obviously *does* reflect a spatial dispersion of effective chain lengths (present at each given instant in time) and consequently also of a time scale distribution. Our data show that the latter spans from the structural relaxation time all the way to about the dielectric Debye time scale. This span entails the mean life time of a given alcohol molecule within an end-to-end hydroxyl chain (termed $\tau_{OH}$ in Ref. 14). This chain was identified as being subject to a process that may be called 'structure diffusion' [32]. The relatively long Debye time scale $\tau_D$ required to render the underlying structures isotropic reflect (i) a molecular property and (ii) an environmental one. For 2E1H with its chain-like supramolecular association the end-to-end vector is roughly collinear to the average molecular dipole moment unlike for 4M3H where due to sterical hindrance of the hydroxyl group ring-like structures dominate, at least below 200 K [16].

However, the long isotropization times leading to $\tau_D$ indicate that structure diffusion pathways exist which are sustained by an environment which provides an only weakly curved tube. While in MAs it is clear that this soft confinement is subject to constant restructuring, it will nevertheless enhance the liquid's rigidity somewhat up to time scales of roughly $\tau_D$. Using a variety of techniques it was found that the supramolecular structures in octanols tend to disintegrate thermally at temperatures near 250 K [28] at which $\tau_D$ is about 100 ns. Thus, our results suggest that high-frequency mechanical probes such as Brillouin scattering experiments should not display a significant signature of the Debye process, in accord with corresponding experimental reports [10].

In summary, we have shown that the Debye process of MAs is not a mere dielectric curiosity. Rather, this process displays the viscoelastic features expected for a short-chain polymer liquid in which the MA's end-to-end contour is stabilized by hydrogen bonds. Our findings open the venue to direct investigations of the particular dynamics of an important class of hydrogen-bonded liquids using not just one single technique, but an entire host of high-resolution experiments such as photon correlation spectroscopy, depolarized light scattering, rheo-NMR, and so on. Likewise, MAs and most likely also peptide-bonded liquids [33] can now be analyzed theoretically in terms of polymer concepts.

**Acknowledgements**

Support of this project by the Deutsche Forschungsgemeinschaft under Grant No. BO1301/8-2 is gratefully acknowledged. The centre for viscous liquid dynamics "Glass and Time" is sponsored by the Danish National Research Foundation's grant No. DNRF61.

**Figures and Captions**

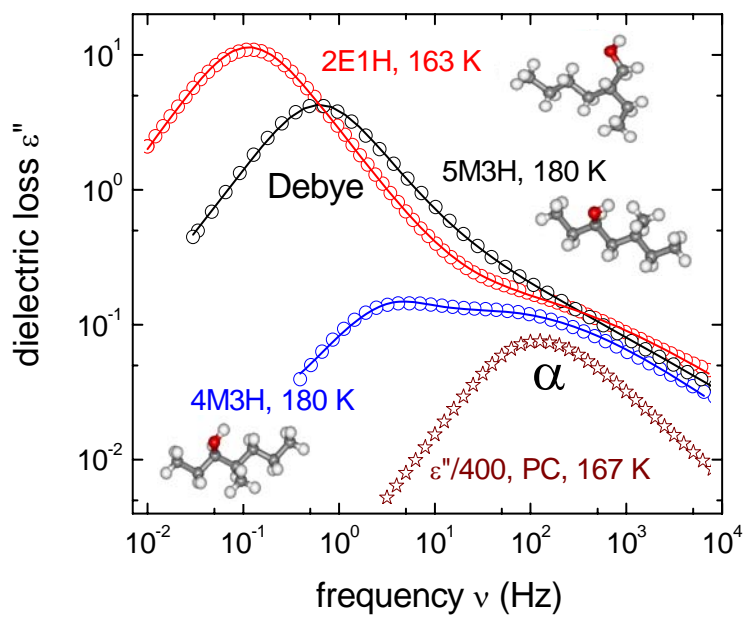

Fig. 1

(Color online) Dielectric loss spectra of the alcohols 4M3H, 5M3H, and 2E1H at temperatures at which the frequency of the α-relaxation is close to 100 Hz for all systems. The strengths of the Debye process of the alcohols differ significantly, whereas the intensity of their α-relaxation is always rather similar. For comparison, the loss spectrum of propylene carbonate (PC) – a typical glass former devoid of hydrogen bonding – is included as indicated.

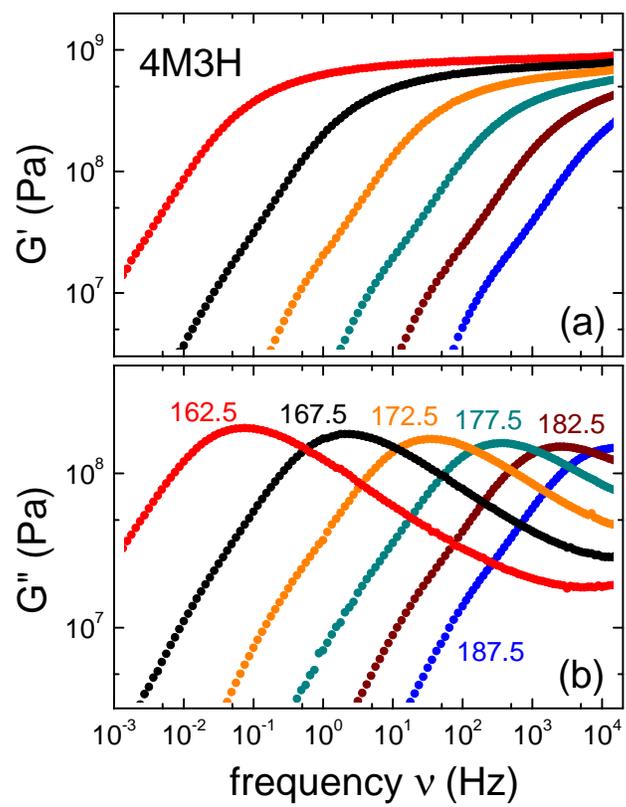

Fig. 2

(Color online) (a) Real part $G'(\nu)$ and (b) imaginary part $G''(\nu)$ of the complex shear modulus of 4M3H for several temperatures (indicated in Kelvin).

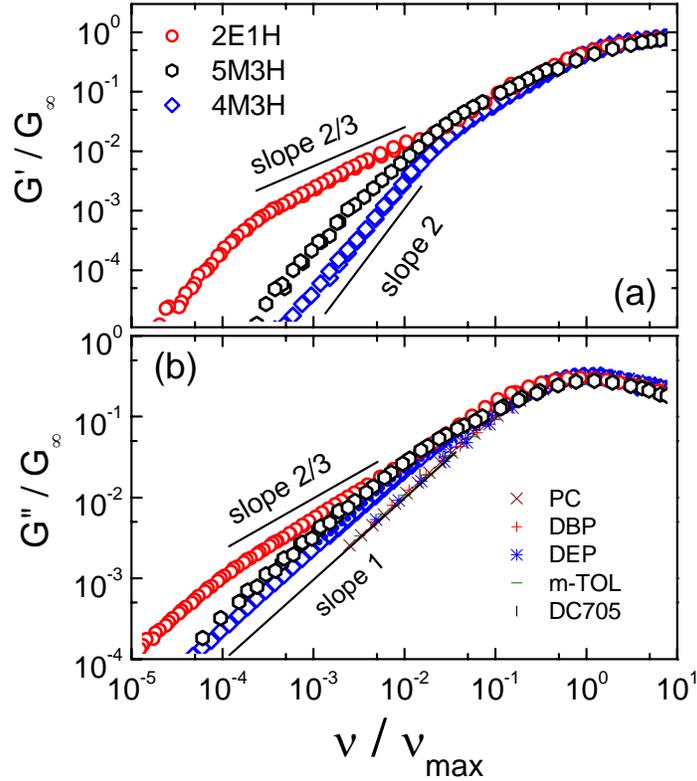

Fig. 3

(Color online) (a) Real and (b) imaginary part of the reduced shear modulus as scaled by the high-frequency shear constant $G_\infty$ ($\approx$ 1 GPa as estimated from the current and earlier [11] broad-band experiments). This plot includes data for the presently studied alcohols, as well as data from Ref. [24] for liquids devoid of a Debye process. These latter liquids also obey time temperature superposition [24] and comprise propylene carbonate (PC), di-butyl-phthalate (DBP), di-ethyl-phthalate (DEP), m-toluidine (TOL), and pentaphenyl-trimethyl-trisiloxane (DC705). The solid lines depict various power laws, see the text for details.

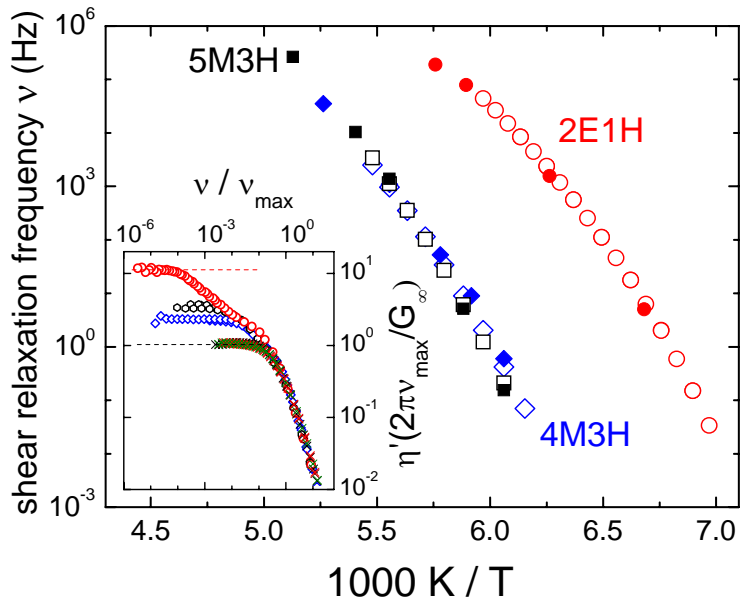

Fig. 4

(Color online) Arrhenius plot of the shear relaxation frequencies. The results from the piezo-ceramic shear transducer experiments (open symbols) agree with those obtained by superimposing the spectra measured with the high-resolution rheological setup (filled symbols). Circles refer to 2E1H, diamonds to 5M3H, and squares to 4M3H. The inset shows $G_\infty$-scaled dynamic shear viscosities of the alcohols and those of small-molecule glass formers devoid of a Debye process (crosses). For the latter liquids $2\pi \nu_{max} \eta'(\nu \to 0)/G_\infty \approx 2\pi \nu_{max} \eta_\alpha/G_\infty \approx 1$ holds, see text, while for the alcohols a sizable departure from this 'simple-liquid' limit is obvious.